\documentstyle[12pt]{article}
\newcommand{\bce}{\begin{center}}
\newcommand{\ece}{\end{center}}
\newcommand{\beq}{\begin{equation}}
\newcommand{\eeq}{\end{equation}}
\newcommand{\bea}{\vspace{0.25cm}\begin{eqnarray}}
\newcommand{\eea}{\end{eqnarray}}

\newcommand{\bsigma}{\mbox{\boldmath $\sigma$}}
\newcommand{\btau}{\mbox{\boldmath $\tau$}}

\newcommand{\bq}{{\bf q}}

\newcommand{\mpi}{m_\pi}

\newcommand{\ba}{\begin{array}}
\newcommand{\ea}{\end{array}}

\newcommand{\ie}{{\sl i.e.~}}
\newcommand{\etal}{{\sl et al.~}}

\newcommand{\ave}[1]{\langle {#1} \rangle}

\def\lsim{\mathrel{\rlap{\lower4pt\hbox{\hskip1pt$\sim$}}
    \raise1pt\hbox{$<$}}}	  %less than or approx. symbol
\def\gsim{\mathrel{\rlap{\lower4pt\hbox{\hskip1pt$\sim$}}
    \raise1pt\hbox{$>$}}}	  %greater than or approx. symbol

\begin{document}
\vspace{1.0in}
\begin{flushright}
{\small February, 1992}
\end{flushright}
\vspace{2.0cm}
\bce
{\large{\bf Model for the Quasifree Polarization-Transfer\\
Measurements in the (p,n) Reaction at 495 MeV}}
\vspace{.55in}

G. E. BROWN

\vspace{0.25cm}
{\it Physics Department\\
State University of New York at Stony Brook\\
Stony Brook, NY 11794-3800}

\vspace{1.0cm}

and J. WAMBACH$\,$\footnote{also at:
Department. of Physics,
University of Illinois at Urbana-Champaign,
Urbana, IL 61801}

\vspace{0.25cm}
{\it
Institut f\"ur Kernphysik\\
Forschungszentrum J\"ulich\\
D-5170 J\"ulich, Fed. Rep. Germany}

\vspace{.35in}
\ece

\vspace{.65in}
\begin{abstract}
The recent $(\vec p,\vec n)$ polarization transfer experiments show
the ratio of spin-longitu-dinal to spin-transverse response functions
$R_L/R_T$, measured at a three-momentum transfer of 1.72 fm$^{-1}$ in
$^{12}$C and $^{40}$Ca, to be essentially unity for small projectile-energy
loss. This can be explained theoretically if the nucleon-nucleon tensor
interaction is zero at this momentum transfer. We can partially achieve
this by introducing a density dependent $\rho$-meson mass $m_\rho^*$ as
well as a nucleon effective mass $m_N^*$. In addition the pion contribution
must further be reduced by softening of the $\pi NN$ form factor.
\end{abstract}

\vspace{.75in}
\begin{flushright}
PACS Indices: 21.30.+y\\
25.40.Ep
\end{flushright}
\baselineskip 20pt
\newpage
\section{Introduction}

Recent experiments on the ($\vec p,\vec n$) reaction by McClelland \etal
\cite{McCl} have strikingly confirmed earlier indications \cite{Care,Rees},
that the ratio of spin-longitudinal to spin-transverse response functions,
$R_L/R_T$, measured at a three-momentum transfer of 1.72 fm$^{-1}$ in
$^{12}$C and $^{40}$Ca is essentially unity for small projectile-energy
losses, decreasing somewhat below unity for energy losses $>$ 70 MeV.
These results put in doubt many theoretical predictions that there should
be a ``softness" in the longitudinal channel due to the pion-exchange tail
of the $NN$ interaction. At momentum transfers such as those selected in
the experiment this would substantially increase the longitudinal response,
$R_L$, above the transverse one. The experimental findings are consistent,
however, with the high-energy Drell-Yan experiments \cite{Alde} which show
no enhancement of the pion cloud in nuclei as compared to that of the free
nucleon.

We propose a model, and carry out a schematic calculation, which connects
the analysis of the quasifree polarization-transfer measurements with the
Drell-Yan experiments \cite{Alde}, and, also, with the measurements of the
charge and current response in quasielastic electron scattering \cite{SoBR}.
By pushing our assumptions somewhat, possibly to an extreme, we are able to
explain the quasifree polarization transfer measurements. After outlining
the underlying ideas we give results of a semi-realistic model calculation
which bears out most of the features of the schematic treatment.

\section{Development}
The quasielastic $(\vec p,\vec n)$ polarization transfer experiments are
interpreted theoretically in terms of a spin-isospin interaction, usually
taken to be (Eq.~(1) of ref.~\cite{McCl})
\bea
V_{\sigma\tau}(q,\omega)=\biggl ({f^2_\pi\over \mpi^2}\biggr )
\biggl \{\biggl ( g'+{q^2\over\omega^2
-(q^2+\mpi^2)}\Gamma_\pi^2(q,\omega)\biggr )
\bsigma_1\cdot\hat \bq\bsigma_2\cdot\hat \bq\nonumber\\
+\biggl (g'+C_\rho{q^2\over \omega^2-(q^2+m^2_\rho)}
\Gamma^2_\rho(q,\omega)\biggr )
\bsigma_1\times\hat \bq\bsigma_2\times\hat \bq
\biggr \}\btau_1\cdot\btau_2.
\eea
It includes pion-and rho-meson exchange and accounts for short-range
correlations via the parameters $g'$ and $C_\rho$. The notation suggests
that the longitudinal response is determined by pion exchange, except for
$g'$, and that the transverse response is governed by $\rho$-meson exchange.
This is misleading. As Baym and Brown \cite{BaBr} made clear, because of
short-range correlations, the $\rho$-meson contributes importantly to $g'$
the longitudinal response is strongly affected. This argument will be
essential for our discussion.

When looking at differences in the interaction which could cause a deviation
of $R_L/R_T$ from unity, it is more transparent to go back to the
decomposition in terms of central and tensor invariants. Ignoring formfactors
for the moment one has
\bea
V_{ten}(q,\omega)=\biggl \{-\biggl ({f_\pi^2\over \mpi^2}\biggr )
\biggl [{\bsigma_1\cdot\bq\bsigma_2\cdot\bq-
{1\over 3}(\bsigma_1\cdot\bsigma_2)q^2\over (q^2+\mpi^2)-\omega^2}\biggr
]\nonumber\\ +\biggl ({f_\rho^2\over m_\rho^2}\biggr )\biggl [
{\bsigma_1\cdot\bq\bsigma_2\cdot\bq-{1\over 3}
(\bsigma_1\cdot\bsigma_2)q^2
\over (q^2+m_\rho^2)-\omega^2}\biggr ]\biggr \}\btau_1\cdot\btau_2  .
\eea
Only this part can cause a difference (see \cite{BaBN} for a review) and
another way is interpreting the experiment is to say that the tensor force
is not active. For our further discussion we adopt as values of coupling
constants
\beq
f_\rho^2/m_\rho^2=2f_\pi^2/\mpi^2,
\label{eq:frho}
\eeq
which is 5 \% larger than the central value of the H\"ohler and Pietarinen
analysis \cite{HoPi}. It should be mentioned that the Bonn Potential
includes the same $\pi$ and $\rho$-exchange components, although
$f^2_\rho/m_\rho^2$ is taken to be $\sim$ 20 \% smaller that in
(\ref{eq:frho}).  The factor of 2 simplifies our model discussions and is
certainly within the H\"ohler and Pietarinen uncertainty. In the context of
meson exchange  this larger factor, can be justified  by the inclusion of
the two-pion continuum \cite{HiSJ} which we shall make more detailed use
of below.

The rough equality for the responses
\beq
R_L(q,\omega)\cong  R_T(q,\omega)
\eeq
for the range $q\sim 1.72-1.75$ fm$^{-1}$ and small $\omega$, with $R_L$
decreasing somewhat below $R_T$ for these $q$ and $\omega > 70$MeV,
as seems to be required consistently by several experiments
\cite{McCl}-\cite{Rees}, can be accomplished by making
$V_{ten}(q,\omega\approx 0) \sim 0$. This implies that
\beq
{f^2_\pi/\mpi^2\over f^2_\rho/m_\rho^2}{(q^2+m^2_\rho)\over
(q^2+\mpi^2)}\cong 1.
\eeq
Since $\mpi^2\ll q^2\ll m_\rho^2$, this simplifies to
\beq
{f^2_\rho/m_\rho^2\over f_\pi^2/\mpi^2}(1+m^2_\rho/6\mpi^2)^{-1}\cong 1
\label{eq:Estim}
\eeq
for $q\sim 1.72$ fm$^{-1}= 2.4 \mpi$. Unfortunately for this simple
explanation, using $\mpi$ and $m_\rho$ from the Particle Data book,
this ratio is 0.4 instead of unity.

One might argue that a number of effects have been left out, but we shall
show by more detailed calculations later that our simple estimate is quite
realistic. In particular, effects of two-body correlations are not large
since the tensor interaction chiefly connects two-nucleon states of relative
zero angular momentum to $L=2$ states and the centrifugal barrier in the
latter suppresses correlation effects . Note that because of this barrier,
effects come chiefly from distances larger than $r\sim\hbar/q$, and the
meson-exchange model should be adequate.

Brown and Rho \cite{BrRo} have suggested that meson masses as well as
the nucleon effective mass should decrease with increasing density
{\it in medium}. Given that \cite{BaBN}
\beq
{f_\rho\over m_\rho}=g_{\rho NN}{(1+\kappa^\rho_V)\over 2m_N}
\eeq
and, in accordance with \cite{BrRo} not scaling $g_{\rho NN}$, we find
that the condition (\ref{eq:Estim}) can be satisfied if
\beq
{m^2_\rho\over m^{*2}_\rho}{m_N^2\over m^{*2}_N}\cong \biggl(
{m_\rho\over m^*_\rho}\biggr )^4{g_A\over g_A^*}=1.
\label{eq:Mroest}
\eeq
In this estimate we included the loop correction to $m^*_N$ found in
the Skyrme model and reproduced in Brown and Rho \cite{BrRo}. Loop
corrections also enter in $g_{\pi NN}$ from which $f_\pi$ derives,
decreasing $g_{\pi NN}$ slightly with increasing density \cite{BrMa}.
But this decrease is small compared with that for $g_A$ and we neglect
it.

The pion mass $\mpi$ which is generated by explicit chiral symmetry
breaking, does not scale as the other masses \cite{BrRo,DeEE,SHSJ};
to a good approximation it is constant over the range of densities
considered here, although from analysis of pionic atoms it is known
that the optical potential is slightly repulsive so that the {\it in medium}
pion mass actually increases slightly with density \cite{SHSJ}. We neglect
the increase in $\mpi$ as well as the decrease in $g_{\pi NN}$ \cite{BrMa}.
Although both effects decrease the {\it in medium} pionic enhancement,
they are small, at most a few percent.

Taking $g_A^*$ to be $\sim$ unity \cite{MRho,OhWa} in nuclei, we find
$(m_\rho/m_\rho^*)^4\cong 2$, giving
\beq
{m^*_\rho\over m_\rho}=0.84.
\label{eq:Mr84}
\eeq
In obtaining this answer we have taken what we consider to be the
minimum possible $g^*_A$ and a rather large $f^2_\rho/m^2_\rho$, but
our values for both cases have been commonly used in the literature.
QCD sum rule calculations \cite{Brow,HaLe} produce\footnote{Hatsuda and Lee
\cite{HaLe} give 0.82 as their central value. Chanfray and Ericson
\cite{ChEr} have shown that exchange current type processes involving the
virtual pion field decrease the quark condensate further by 10-20 \% over
that of Hatsuda and Lee. Including this correction will bring
$m^*_\rho(\rho_0)/m_\rho$ down to $\sim$ 0.8.}
\beq
{m^*_\rho(\rho_0)\over m_\rho}\cong 0.8.
\eeq
which is not so different from our estimate.

\section{Numerical Results}

For a more quantitative analysis  we use a spin-isospin interaction based
on the recent work by Hippchen \etal  \cite{HiSJ} which, in addition to the
$\rho$ meson, includes effects of the $\pi\pi$ continuum. This interaction
(OBEPH)  is consistent with "strong $\rho$ " coupling  as obtained from
the H\"ohler-Pietarinen analysis . Effects of short-range correlations
are included  by multiplying  $V_{\sigma\tau}$  by a correlation function
$f(r)=1-j_0(r/r_c)$ with $r_c=0.19$ fm \cite{HiSJ} . This value for
$r_c$ has been extracted from a fit to the scattering observables over a
wide range of energies. The resulting static spin-longitudinal and
spin-transverse interactions are displayed as the dashed lines in Fig.~1.
They show the familiar $q$-dependence. It is instructive the extract the
Fermi-Liquid parameter $g'$ and the $C_\rho$ factor from this interaction,
which are used in phenomenological parameterizations such as Eq.~(1).
The particular form used in Eq.~1 is not very suitable , however, since
it results in a strong $q$-dependence of these parameters. A better choice
\cite{Chan} is
\bea
V_{\sigma\tau}(q,\omega)=\biggl ({f^2_\pi\over \mpi^2}\biggr )
\Gamma_\pi^2(q,\omega)
\biggl \{\biggl ( g'+{q^2\over\omega^2
-(q^2+\mpi^2)}\biggr )
\bsigma_1\cdot\hat \bq\bsigma_2\cdot\hat \bq\nonumber\\
+\biggl (\Gamma_\pi^2(q,\omega)g'+C_\rho{q^2\over \omega^2-(q^2+m^2_\rho)}
\Gamma^2_\rho(q,\omega)\biggr )
\bsigma_1\times\hat \bq\bsigma_2\times\hat \bq
\biggr \}\btau_1\cdot\btau_2.
\eea
Comparing the static limits of the OBEPH potential and $V(q,0)$ makes it
possible
to extract then $g'$ and $C_\rho$ . The results for $g'$ are given by the
dashed lines in the lower part of Fig.~1.  The momentum dependence is weak
and the values are smaller than those extracted from Gamow-Teller
systematics ($g'\sim 0.7-0.8$). The parameter $C_\rho=3.65$ is also found
to be weakly $q$ dependent but it is significantly larger than what is
typically used for "strong  $\rho$" coupling ($C_\rho\sim 2$), as to be
expected from the inclusion of  the $\pi\pi$ continuum \cite{HiSJ}.

With medium-modified hadron masses the only parameter which enters the
spin-isospin interaction is  $m_\rho^*$. This is easily incorporated in
the OBEPH potential.  One can then treat $m_\rho^*/m_\rho$ as a parameter
so as to force the tensor interaction $V_{ten}$ zero at the relevant
momentum transfer, as discussed in the previous section. It turns out
that the resulting value is close to $m_\rho^*(\rho_0)/m_\rho=0.8$,
constituent with QCD sum rule values and our schematic calculation.
The resulting interactions are shown as the full lines in Fig.~1.
In both, $V_L$ and $V_T$, we find a strong increase in the repulsion with g'
going from $\sim$ 0.6 to a value near unity (lower part of Fig.~1). At the
same time the two interactions become weak and nearly equal near $q_0=1.72$
fm$^{-1}$, the momentum transfer of the experiment. More quantitatively
this is illustrated by the nuclear matter response functions and the ratio
$R_L/R_T$, both at $\rho_0$, which are displayed in Fig.~2.
As is well known, without medium modifications, $R_L$ and $R_T$ are quite
different (dashed lines in the upper part of Fig.~2) and the ratio is
strongly energy dependent (dashed line in lower part). A dropping
$\rho$-meson mass wipes out this effect (full line) as experiment would
suggest. At the same time, $R_L$ and $R_T$ are rather close to the
free response.

While a nuclear matter calculation with dropping $\rho$ mass can explain
the LAMPF measurement one has to bear in mind the surface nature of the
$(p,n)$ reaction. Without performing a sophisticated finite-nucleus DWBA
calculation \cite{Ichi} we can still get some realistic estimate for the
finite nucleus effects by using a semi-classical description of the
response functions and treating the spin-isospin interaction in a local
density approximation (LDA), \ie including a density dependence for
$m^*_\rho/m_\rho$. Following the work of ref.~\cite{Chan} we obtain the
finite-nucleus response functions as
\beq
R^A_{L,T}(q,\omega)=\int d^3 r R^{NM}_{L,T}(q,\omega;\rho(r ))F(r)
\eeq
where $R^{NM}_{L,T}(q,\omega,\rho)$ are the {\it local} nuclear matter
response functions, including longitudinal-transverse mixing through the
surface (see \cite{Chan} for details), and $F(r)$ is a weight factor
derived from the projectile distortion function. It can be evaluated
reliably in the Eikonal approximation.  Assuming a linear density dependence
for the $\rho$-meson mass as $m^*_\rho/m_\rho=1-c\rho/\rho_0$, as suggested
by available QCD sum rule calculations, the results for $^{40}$Ca are shown
in Fig.~3. The upper part displays $R_L$ and $R_T$ with (full lines) and
without (dashed line) dropping mass. As for nuclear matter , their
differences are drastically reduced and the response functions  are close
the free response (dashed-dotted line).  In the ratio the effect is not as
dramatic, however . Nonetheless, there is a significant improvement for
lower energies.

We have also included effects from coupling to isobar-hole
excitations which
are known to increase the tensor interaction in the nuclear medium.
When dropping the $\rho$-meson mass these effects are negligible, however,
because of the smallness of $V_{L,T}$ at $q=1.72$ fm$^{-1}$.

\section{Discussion}
We have investigated a model with dropping hadron masses to explain the
equality of the spin-longitudinal and -transverse responses, as inferred
 from
recent quasifree polarization-transfer $(\vec p,\vec n)$ measurements at
LAMPF. A decrease of the $\rho$ meson mass with density leads to an
increase
in its tensor interaction which alters the momentum dependence of the
longitudinal and transverse interactions. In nuclear matter and
semi-classical
finite-nucleus calculations,  a value of $m_\rho^*/m_\rho=0.8$ explains the
data. Such a value seems low, given the surface nature of the $(p,n)$
reaction. Using a Glauber estimate, we find that the nucleus is probed at
$\sim$ 0.6 $\rho_0$ which yields $\ave{m^*_\rho/m_\rho}=0.88$, when using a
linear density dependence. With such a value  $R_L/R_T$ is still reduced but
not enough to explain the data. A more detailed LDA calculation confirms
this accurately. On the other hand, an effective $\rho$-meson mass of 0.88
fits well with values deduced from other experiments. For instance, in an
analysis of $(e,e')$ and $(p,p')$ scattering data at 318 MeV on the
stretched $12_{1,2}^-$ and $14^-$ states in $^{208}$Pb \cite{HiLS} agreement
between the electron and proton quenching factors can be found for effective
masses
\beq
\ave{m^*_\rho\over m_\rho}=0.79-0.86 .
\eeq
Similarly, from studies of the $(\vec p,\vec p')$ polarization transfer
at $E_p=200$ MeV in $^{16}$O, Stephenson and Tostvin \cite{StTo} find
that lowering $m^*_\rho$ to
\beq
\ave{m^*_\rho/m_\rho}=0.94\pm 0.2
\eeq
changes the effective isovector spin interaction so as to significantly
improve the fit to the data. In $^{10}$B $(\vec p,\vec p')$, at the same
energy, Baghaei \etal \cite{Bagh} find that decreasing
$\ave{m^*_\rho/m_\rho}$ to 0.9 substantially improves their fit to the data.

It is clear that lowering $\ave{m_\rho^*/m_\rho}$ to the $\sim$ 0.84 of
Eq.~(\ref{eq:Mr84}) would improve our fit to the data. Alternatively we can
seek to cut down the pion exchange contribution relative to that of the
$\rho$-exchange, chiefly at the relatively large momentum transfer of
$q=1.72$ fm$^{-1}$ of the polarization transfer experiment. This can be
achieved by changing the pion form factor $\Gamma_\pi(q,\omega)$. Whereas a
monopole form factor with a cut off $\Lambda_\pi=2$ GeV was used in the
OBEPH of Ref.~\cite{HiSJ} Huang \etal \cite{HwSB} find that $\Lambda_\pi$
should be lowered to $\sim$ 950 MeV in order to fit the sea quark
distribution in deep inelastic lepton scattering from hadrons ( others
\cite{Thom,FrMS} have argued for an even lower value). More recently
a microscopic model using a $\pi\rho$ vertex and including
rescattering contributions has confirmed these findings \cite{JaHS}.
Such a low
form factor, of course, cannot describe the deuteron properties. To
reconcile this difficulty Holinde and Thomas have introduced a heavy
point-like "pion" (the $\pi$') of mass 1.3 GeV \cite{HoTh}.
By adjusting the $\pi'NN$ coupling constant good fits to the phase
shifts and the deuteron properties could be obtained. This picture,
in conjunction with the $\pi\rho$ vertex model \cite{JaHS}, then allows
to study
medium modifications of the $\pi NN$ form factor. Such modifications are
again produced by a dropping $\rho$ mass. Preliminary calculations
\cite{Jans} indicate a significant softening, chiefly from the rescattering
term. The "effective" $\pi NN$ cut off in the OBEPH potential drops
from 2 GeV to about 1 GeV. Changing $\Lambda_\pi$ to 1 GeV leads to
the dashed-dotted curve in Fig.~3, which is quite close to the data.

We conclude that inclusion of the {\it medium dependence of the $\rho$-meson
mass and of the nucleon effective mass} significantly reduces the
isovector part of the tensor interaction and removes much of the
discrepancy between the ratio of spin-longitudinal and spin-transverse
response functions. Agreement with experiment further improves by a
softening of pion-exchange form factor. In microscopic models of the
$\pi NN$ vertex such a softening is caused by the same physical mechanism.
Our fits deteriorate for
higher values of $\omega$, however, and new physics, which we have not
considered, may enter there.

The medium dependence of the $\rho$ mass leads to a very different
momentum dependence of the spin-isospin interaction as that in free space
(see Fig.~1). This implies  a different $q$-dependence of the ratio
$R_L/R_T$. Our predictions are given in Fig.~4. At
$q$=1.1 fm$^{-1}$, currently measured at LAMPF \cite{Mcl1}, one should
observe a significant enhancement at low $\omega$ (dashed line)
while at $q$-tranfers $>$ 2 fm$^{-1}$ the ratio should be strongly
suppressed since $V_T$ becomes attractive (dashed-dotted line). We
therefore consider measurements at different $q$-tranfers crucial to
our model.

\vspace{2.0cm}

\bce
{\bf Acknowledgement}
\ece

\noindent
We thank G. Jan\ss en for calculations of the in-medium $\pi NN$
form factor and J. McClelland for informations on the data.
This work was supported in part by DOE grant
DE-FG 0188ER40388 and NSF grant PHY-89-21025.

\newpage
\vspace{.25in}
\parindent=.0cm            %align refs

\newpage
\begin{center}
{\bf Figure Captions}
\end{center}

\noindent Fig.~1: upper part: the spin-longitudinal and spin-transverse
free
OBEPH interactions (dashed lines) and the medium-modified interactions
for
$m^*_\rho/m_\rho=0.8$ (full lines). lower part: the Fermi-liquid parameter
$g'(q)$ in free space (dashed line) and in medium (full line).

\vspace{0.20in}

\noindent Fig.~2: upper part: the nuclear matter spin-longitudinal and
spin-transverse response functions as a function of energy $\omega$ and
at fixed momentum transfer $q_0=1.72$ fm$^{-1}$. The dashed lines use
the free OBEPH interaction while the full lines result from dropping the
$\rho$-meson mass. The dashed-dotted line denotes the free Fermi gas
response.
lower part: the ratio $R_L/R_T$  from the OBEPH interaction (dashed) and
with dropping mass (full). The data were taken from ref.~\cite{McCl}.
\vspace{0.20in}

\noindent Fig.~3: same as Fig.~2 but for $^{40}$Ca. The dashed-dotted line
gives results for a cut off $\Lambda_\pi=1$ GeV.
\vspace{0.20in}

\noindent Fig.~4: the ratio $R_L/R_T$ for several momentum transfers as
predicted by our model. A "soft" cut off $\Lambda_\pi=1$ GeV has been
used.


\begin{thebibliography}{99}
\bibitem{McCl} J. B. McClelland at al., Phy. Rev. Lett. {\bf 69},
582 (1992).
\bibitem{Care} T. A. Carey at al., Phys. Rev. Lett. {\bf 53}, 144 (1984).
\bibitem{Rees} L. B. Rees at al., Phys. Rev., {\bf C34}, 627 (1986).
\bibitem{Alde} D. M. Alde et al., Phys. Rev. Lett., {\bf 64}, 2479 (1990).
\bibitem{SoBR} M. Soyeur, G. E. Brown and M. Rho, Nucl. Phys. {\bf A},
 to be published;\\
G. E. Brown, M. Rho and M. Soyeur, Proc. Int. Conf. on Nucl. Phys.,
Wiesbaden, 27. July-1. Aug.,1992, Nucl. Phys. {\bf A}, to be published;
 Acta Physica Polonica, to be published.

\bibitem{BaBr} G. Baym and G. E. Brown, Nucl. Phys. {\bf A247}, 395
(1975).
\bibitem{BaBN} S.-O. B\"ackman, G. E. Brown and J. A. Niskanen,
Phys. Rept. {\bf 124}, 1 (1985).
\bibitem{HoPi} G. H\"ohler and E. Pietarinen, Nucl. Phys. {\bf B95},
210 (1975).
\bibitem{HiSJ} T. Hippchen, J. Speth and  M. B. Johnson, Phys. Rev.
{\bf C40}, 1316 (1989).
\bibitem{BrRo} G. E. Brown and M. Rho, Phys. Rev. Lett. {\bf 66},
2720 (1992).
\bibitem{DeEE} J. Delorme, M. Ericson and T. E. O. Ericson,
Phys. Lett. {\bf 291B}, 379
(1992).
\bibitem{SHSJ} C. Sch\"utz, K. Holinde, J. Speth and M. B. Johnson,
KFA preprint.
\bibitem{BrMa} G. E. Brown and R. Machleidt, preprint SUNY-NTG-92-30.
\bibitem{MRho} M. Rho, Nucl. Phys. {\bf A231}, 443 (1974).
\bibitem{OhWa} K. Ohta and M. Wakamatsu, Nucl. Phys. {\bf A234}, 445
(1975).
\bibitem{Brow} G. E. Brown, Nucl.Phys. {\bf A231}, 397c (1991).
\bibitem{HaLe} T. Hatsuda and S. H. Lee, Phys. Rev. {\bf C46}, R34
(1992).
\bibitem{ChEr} G. Chanfray and M. Ericson, Nucl. Phys. {\bf A},
(1993) to be published.
\bibitem{Ichi}  M. Ichimura, K. Kawahigashi, T. S. Jorgensen and
C. Gaarde, Phys. Rev, {\bf C39}, 1446 (1989).
\bibitem{Chan}  G. Chanfray, Nucl. Phys. {\bf A474}, 114 (1987).
\bibitem{BrLL}  G. E. Brown, Z. B. Li and K. F. Liu, Nucl. Phys.
{\bf A}, to be published.
\bibitem{HiLS} N. M. Hintz, A. M. Lallena and A. Sethi, Phys. Rev.
 {\bf c45}, 1098 (1992).
\bibitem{StTo} E. J. Stephenson and J. A. Tostevin, in
{\it Spin and Isospin in Nuclear Interactions}, ed. by S. W.
Wissink et al. (Plenum, New York, 1991) p. 281.
\bibitem{Bagh} H. Baghaie et al., Phys. Rev. Lett. {\bf 69}, 2054 (1992).
\bibitem{HwSB} W-Y. Hwang, J. Speth and G. E. Brown, Z. Phys.
{\bf A339}, 383 (1991).
\bibitem{Thom} A. W. Thomas, Phys. Lett, {126B}, 97 (1983).
\bibitem{FrMS} L. L. Frankfurt, L. Mankiewicz and M. I. Strikman,
Z. Phys. {\bf A334}, 343 (1989).
\bibitem{JaHS} G. Jan\ss en, K. Holinde and J. Speth, to be published.
\bibitem{HoTh} K. Holinde and A. W. Thomas, Phys. Rev. {\bf C42}, R1195
(1992).
\bibitem{Jans} G. Jan\ss en, private communication.
\bibitem{Mcl1} J. B. McClelland, private communication.
\end{thebibliography}
\end{document}